\documentclass[prl,%
 reprint,
 amsmath,amssymb,
 aps,
]{revtex4-2}

\usepackage{graphicx}
\usepackage{dcolumn}
\usepackage{bm}

\usepackage[inkscapelatex=false]{svg}
\usepackage{siunitx}
\usepackage{textcomp,gensymb} 
\usepackage{cleveref}
\usepackage{multirow}
\usepackage{amsfonts,amssymb}
\usepackage{tikz,pgfplots,pgfmath}
\usepackage{amsmath}
\usepackage{amssymb}
\usepackage[english]{babel}
\usepackage{tikz,pgfplots,pgfmath}
\pgfplotsset{compat=1.18}
\usepackage{makecell}
\usepackage{array}
\usepackage[utf8]{inputenc} 
\usepackage{url}            

\begin{document}

\preprint{APS/123-QED} 

\title{Simultaneous PW-scale laser driven MeV X-ray and neutron beam characterization for dual probing 
capability}


\author{I. Cohen$^{1,2}$}
\email{Itamar.oarb@gmail.com}
\thanks{Corresponding author: I. Cohen, Technion Israel Institute of Technology, Faculty of Physics, Technion City, Haifa 3200003, Israel}

\author{W. Yao$^{2,3}$}
\author{N. Mirkovic$^{2}$}
\author{P. Antici$^{4}$}
\author{G. Augé$^{5}$}
\author{P.-G. Bleotu$^{6}$}
\author{T. Catabi$^{7}$}
\author{S. N. Chen$^{6}$}
\author{A. Ciardi$^{3}$}
\author{F.P. Condamine$^{8,15}$}
\author{E. d’Humières$^{9}$}
\author{Q. Ducasse$^{10}$}
\author{G. Fauvel$^{8,9}$}
\author{R. Gambicchia$^{10}$}
\author{G. Giubega$^{6}$}
\author{L. Gremillet$^{11,12}$}
\author{M. Gugiu$^{6}$}
\author{V. Iancu$^{6}$}
\author{R. Lelièvre$^{10}$}
\author{L. T. Mix$^{13}$}
\author{Y. Ristic$^{10}$}
\author{D. Sangwan$^{6}$}
\author{M. Sheats$^{13}$}
\author{F. Trompier$^{10}$}
\author{L. Tudor$^{6}$}
\author{S. Turiel$^{1}$}
\author{G. Verstraeten$^{14}$}
\author{T. Vinchon$^{10}$}
\author{I. Pomerantz$^{7}$}
\author{O. Tesileanu$^{6}$}
\author{J. Fuchs$^{1,2}$}

\affiliation{$^1$ Technion Israel Institute of Technology, Faculty of Physics, Technion City, Haifa 3200003, Israel}

\affiliation{$^2$ LULI - CNRS, CEA, UPMC Univ Paris 06 : Sorbonne Université, Ecole Polytechnique, Institut Polytechnique de Paris - F-91128 Palaiseau cedex, France}

\affiliation{$^3$ Sorbonne Université, Observatoire de Paris, PSL Research University, LUX, CNRS, 75005 Paris, France}

\affiliation{$^4$ INRS-EMT, 1650 boul. Lionel-Boulet, Varennes, QC, J3X 1S2, Canada}

\affiliation{$^5$ TRACTEBEL ENGINEERING S.A., 34 boulevard Simon Bolivar, 1000 Bruxelles, Belgium}

\affiliation{$^6$ Extreme Light Infrastructure - Nuclear Physics, "Horia Hulubei" National Institute for R\&D in Physics and Nuclear Engineering, 30 Reactorului Street, 077125 Magurele, Romania}

\affiliation{$^7$ The School of Physics and Astronomy, Tel Aviv University, Tel-Aviv, 6997801, Israel}

\affiliation{$^8$ The Extreme Light Infrastructure, ELI Beamlines, Za Radnici 835, 252 41 Dolni Brezany, Czechia}

\affiliation{$^9$ CELIA, University of Bordeaux, CNRS, CEA, UMR 5107, F-33405 Talence, France}

\affiliation{$^{10}$ Laboratoire de Micro-irradiation, de Métrologie et de Dosimétrie des Neutrons, PSE-SANTE/SDOS, Autorité de Sûreté Nucléaire et de Radioprotection (ASNR), 13115 Saint-Paul-Lez-Durance, France}

\affiliation{$^{11}$ CEA, DAM, DIF, F-91297 Arpajon, France}

\affiliation{$^{12}$ Université Paris-Saclay, CEA, LMCE, F-91680 Bruyères-le-Châtel, France}

\affiliation{$^{13}$ LANL, PO Box 1663, Los Alamos, New Mexico 87545, USA}

\affiliation{$^{14}$ Vrije Universiteit Brussel, Brussels 1000, Belgium}

\affiliation{$^{15}$ GenF, 2 Avenue Gay Lussac 78990 Elancourt, France}

\date{\today}

\begin{abstract}
Laser-driven, high-brilliance secondary sources (electrons, ions, neutrons, X-rays) open new perspectives for compact material probing and imaging of high-speed events. A key advantage is their ability to perform multiplexed probing, as these sources are generated simultaneously in a single shot using a single laser beam. Here, we report the first quantitative measurements of photon spectra (0.1--100 MeV) and angular distributions in the petawatt interaction regime, using an ultra-intense ($>10^{21}\,\rm W/cm^2$), ultra-short (24~fs) laser pulse. These results are complemented by the characterization of simultaneously produced MeV neutrons. We discuss how these neutrons, once moderated, can enable in-depth material identification via resonance transmission analysis. This work highlights the potential of compact, ultrashort-pulse PW lasers for dual neutron and X-ray radiography of dense materials.
\end{abstract}

\maketitle


\section{Introduction}\label{introduction}

X-ray and neutron sources generated by ultra-intense lasers 
have made great strides in recent years with the development of laser technology. There are many ongoing efforts worldwide to produce laser-driven, high-brightness neutron beams and to examine their potential use in radiography and material probing applications~\cite{yogo2023laser, Kishon2019laser, yao2025characterization, Gnther2022}. When an ultra-intense laser pulses impinges onto a thin solid foil, a dense bunch of relativistic ($\sim \rm MeV$) electrons is first accelerated \cite{Beg1997, Haines2009, Chen2009, Kluge2018,  Rusby2024, Liseykina2015}, 
leading to the following secondary photon and particle sources.

First, the fast electrons radiate within the target material, resulting in the emission of an 
intense flash of X-rays and $\gamma$ rays in an energy range comparable with that of the electrons. This emission can be produced by at least two concurrent processes \cite{Faenov2015, Martinez2020synchrotron, Liang2023}: (i) synchrotron radiation in the strong laser or self-induced fields \cite{Nerush2014}, and (ii) Bremsstrahlung emissions due to collisions with the target nuclei, the latter being especially effective in thick high-$Z$ targets \cite{vyskovcil2018simulations, Strehlow2024}.

Second, the space-charge fields induced by the fast electrons around the target drive a population of fast ions (mostly protons) from the target surfaces, the most energetic ones originating from the target rear \cite{Macchi2013,Daido2012}. These protons, injected in a ``catcher'' secondary target, can induce neutron-generating nuclear reactions \cite{kleinschmidt2018, lelievre2024comprehensive, lelievre2024high}, with the neutron yield scaling strongly with the incident laser energy \cite{lelievre2024high}. A record yield of $10^{10}$ neutrons/sr per laser shot has been reported \cite{kleinschmidt2018}.

Alternatively, neutrons can also be produced, still in the ``pitcher-catcher'' scheme \cite{Willingale2011}, but employing laser-driven wakefield accelerated electrons as the primary source, which are then converted into hard photons and neutrons through suitable ($\gamma$,n) reactions in a high-Z converter \cite{Vallires2025}. Another method,  to also generate the neutrons by  energetic (tens of MeV) X-rays and exploiting the same photonuclear reactions, is to directly laser-irradiate thick, high-$Z$ targets  \cite{Hatchett2000,Pomerantz2014, Liang2023, Cohen2024}. Note that target engineering, such as placing low-density foams ahead of the solid target to increase the laser intensity and its conversion into electrons, would further boost the subsequent yield of X-rays and neutrons \cite{Gnther2022, Horny2022-tn}. 

However, beyond the pioneer work of Hatchett et al. in 2000 \cite{Hatchett2000}, the characterization of the X-rays emitted from laser-driven foil targets has been comparatively much less investigated, especially compared to X-ray sources generated by wakefield-driven electrons in gas targets \cite{Albert2023}. This paper addresses this gap by providing the first comprehensive characterization of X-ray emission--including energy spectra, angular distributions, and conversion efficiencies--from solid foils irradiated at the petawatt scale in the ultra-relativistic regime enabled by ultra-short pulses (here 24 fs).

In addition, we present a detailed analysis of the simultaneously generated electron, proton, and neutron distributions, allowing for a thorough assessment of the PW-laser-driven radiation. Finally, we evaluate the utility of this dual-source (neutron and X-ray) system for material probing, as discussed by \cite{Brenner2015} and already demonstrated by \cite{Yogo2021,Wei2022-mb,Armstrong2024,Treffert2024}. We specifically demonstrate that the proven capabilities of MeV X-ray tomography \cite{Hollinger:25} can be effectively complemented by neutron resonance transmission analysis (NRTA) \cite{Postma2017-yn} for the non-destructive inspection of nuclear-related dense materials, provided the MeV neutrons are appropriately moderated.


The paper is organized as follows. (I) We first present the experimental setup and the laser characteristics. (II) We detail the characterization of the fast (MeV) electrons produced by the laser-solid interactions. (III) We detail the characterization of the X-rays produced by the electrons within the target. (IV) We present the characteristics of the protons accelerated by the electrons. (V) We overview the features of the neutrons that are produced by the protons in a downstream converter. (VI) Finally, we discuss the prospects of using such laser-driven dual beams for material probing. (VII) A discussion concludes the paper.

\section{Experiment}
\label{Experiment}

\subsection{Experimental setup}
\label{Sec_Experimental}

\begin{figure}[hbtp]
\centering
\includegraphics[width=0.45\textwidth]{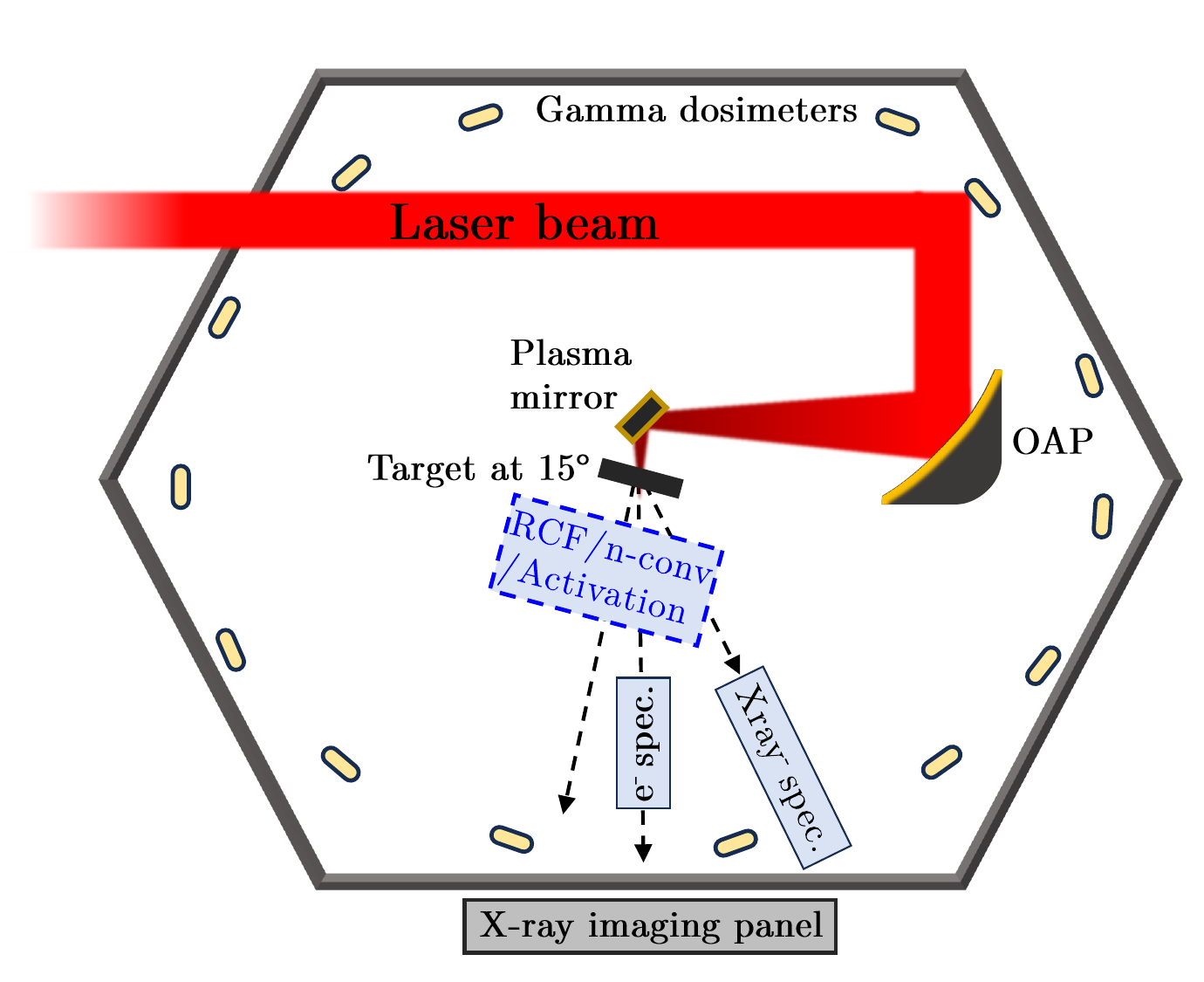}
\caption{Sketch of the experimental setup and of the arrangement of the diagnostics (top view, see text for details). OAP stands for off-axis-parabola, the focusing optics used to focus the laser onto the target.  The target was tilted by 15$^{\circ}$ compared to the laser axis, in order to minimize back-reflection of laser light into the incident laser. Further, a reflective (Au-coated) plasma mirror was used close to the focus (at 30 mm from the target). Its intend was not to improve the laser contrast, as usually done when using anti-reflection coated plasma mirrors \cite{Doumy2004-lf}, but merely to lower back-reflection from the target into the laser chain. In the blue dashed box is a set of diagnostic that can be translated in the beam of secondary sources generated by the laser-target interaction. RCF films \cite{bolton2014instrumentation} serve to analyze the produced proton beam. A neutron converter (LiF) can be also placed in the path to generate neutrons from (p,n) reactions. These neutrons are then analyzed with a set of nuclear activation detectors. The X-ray imaging panel was located at 1.15~m from the target.
}
\label{exp_setup}
\end{figure} 

\begin{figure}[hbtp]
\centering
\includegraphics[width=0.45\textwidth]{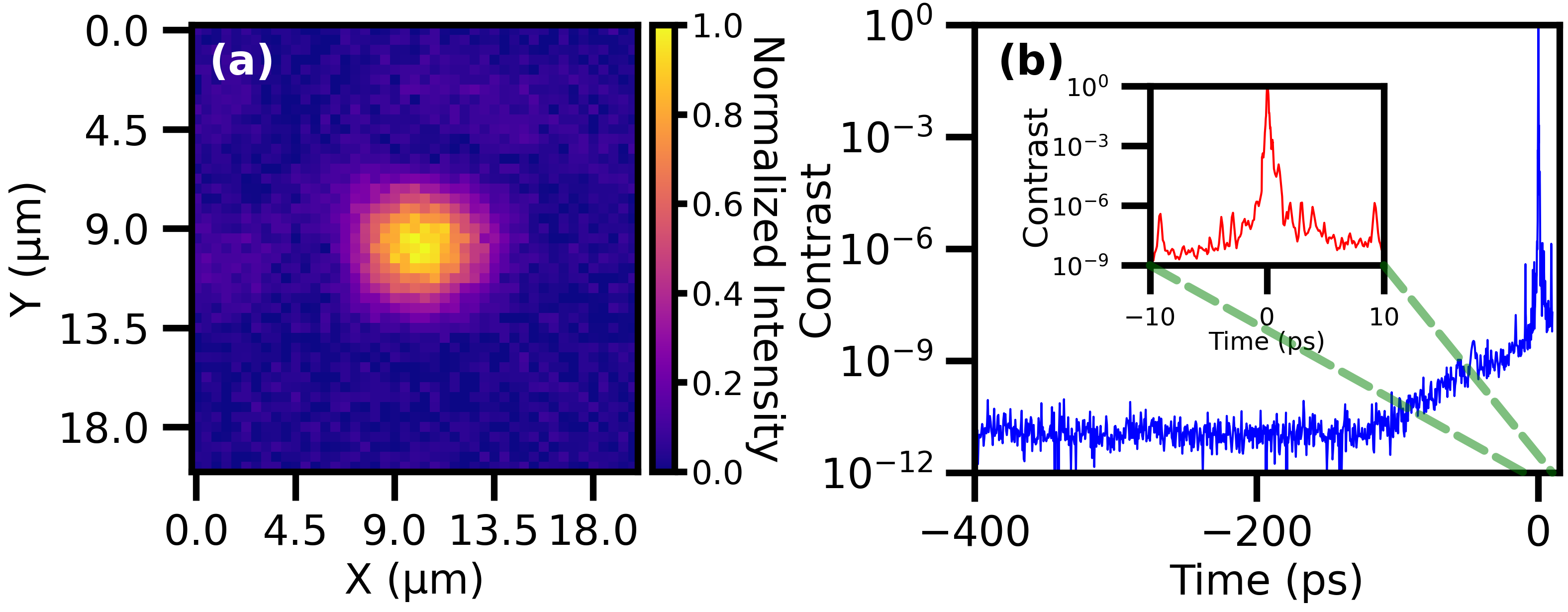}
\caption{(a) Laser focal spot, as measured at low power by a high-resolution, high-aperture in-vacuum microscope objective linked to a CCD camera. (b) Laser pulse contrast measurements. Time 0 corresponds to the peak of the laser pulse. The scale starts at -400 ps to show the laser-prepulse that initiate plasma ablation and expansion prior to the peak of the pulse. The laser temporal shape within $\pm$ \SI{10}{\pico\second} of the peak of the laser pulse is shown in the inset of (b). The pulse duration was measured with a single-shot FROG device \cite{FROG_1997}.
}
\label{laser}
\end{figure}

The experiment was performed using the 1~PW beamline of the ELI-NP laser facility \cite{Cernaianu2025,Radier2022} (see Fig.~\ref{exp_setup} for the setup). The laser delivered an average energy of $\sim 15\,\rm J$ on target, with a central wavelength of $0.810\,\rm \mu m$. The beam was focused to a $4.5 \ \rm \mu m$ diameter (FWHM) spot using an $f/3.7$ off-axis parabola [Fig.~\ref{laser}(a)]. It had a Gaussian temporal profile with a 24~fs FWHM duration [Fig.~\ref{laser}(b)], resulting in a peak on-target intensity of about $3 \times 10^{21}\,\rm W\,cm^{-2}$.

Although the laser system supported a 1~Hz repetition rate, shots were performed on demand due to the time required to align each solid target. A new target was positioned at the focus for each shot, as the laser interaction transformed the target into plasma, effectively destroying it. The positioning precision along the focal axis was $10\,\rm \mu m$, well within the laser's depth of focus ($\sim 50\,\rm \mu m$). 


\subsection{Electron measurements}\label{Sec_Electrons}

The laser-matter interaction at the target surface instantaneously accelerates electrons to MeV energies. Because the target thickness is much smaller than the electron stopping length, these electrons can fly into the vacuum; this builds up a positive charge within the target, creating an electrostatic potential that retains most electrons and causes them to recirculate across the target \cite{Green2018-bo}. The small fraction of fast electrons with sufficient energy to escape this target potential \cite{Cottrill2010, Rusby2019, Link2011} was collected by a magnetic spectrometer positioned downstream along the laser axis (see Fig.~\ref{exp_setup}), where the emission is expected to be highest \cite{Rusby2015}.

The spectrometer had a design similar to that described in Ref.~\cite{Pomerantz2014}. It was 160~mm long and its entrance was 470~mm away from the target. The spectrometer entrance was shielded by 2~cm of lead containing a 5~mm diameter pinhole. The vertical distance from the pinhole to the detector plane was 10~mm. Electron trajectories within the spectrometer were simulated using the known spatial distribution of the magnetic field (peaking at 0.35~T at the center). 

\begin{figure}[hbtp]
\centering
\includegraphics[trim={0cm 0cm 0cm 0cm},clip, width=0.45\textwidth]{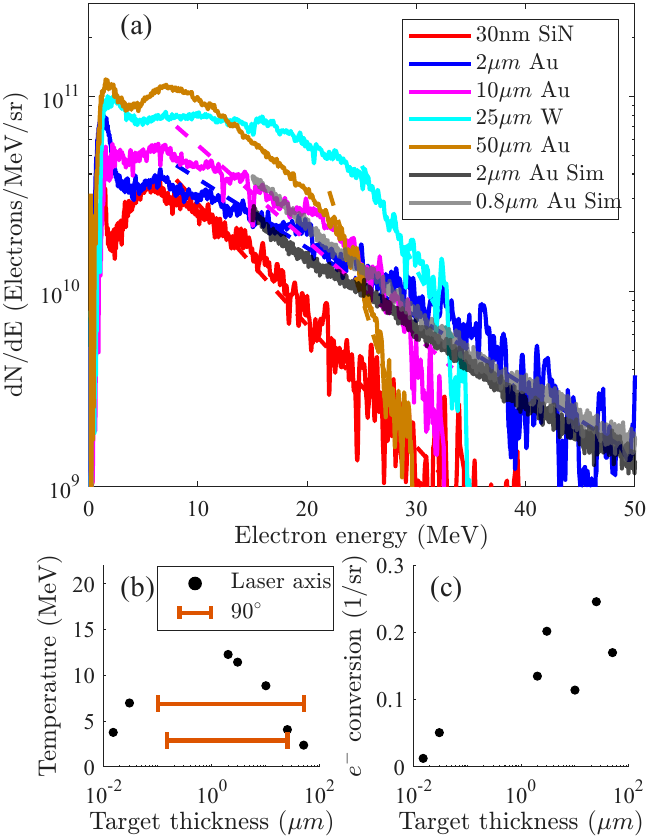}
\caption{(a) Measured and simulated electron spectra from different solid foil target materials and thicknesses. The electrons are measured using absolutely calibrated TR-type imaging plates \cite{Rabhi2016-dk}. The lower limit of the graph corresponds to the noise level on the detector. For each shot, the temperature inferred from fitting the spectrum tail to an exponential, and the conversion efficiency are calculated, as shown respectively in panels (b) and (c).
}
\label{e-spec}
\end{figure}

Electron spectra detected for various target thicknesses and materials are presented in Fig.~\ref{e-spec}(a). Effective electron temperatures were determined by fitting the spectral slope [dashed lines in Fig.~\ref{e-spec}(a)], while the laser-to-electron conversion efficiency was calculated via numerical integration of the measured spectra, as shown in Fig.~\ref{e-spec}(b-c).

The data indicate that the conversion efficiency increases with target thickness, while the electron temperature peaks at a thickness of $2\,\rm \mu m$ Au and decreases for thicker targets. This suggests two distinct interaction regimes: a first regime for targets thinner than $2\,\rm \mu m$, where the entire target is affected by the leading edge of the laser pulse, and a second regime where increased target thickness reduces the electron refluxing,leading to a lower effective temperature \cite{CHEN2005, Buffechoux2010, CompantLaFontaine2013, Huang2022}.

\subsection{Proton measurements}\label{Sec_Protons}

\subsubsection{Proton energy and angular distributions.}

\begin{figure}[hbtp]
\centering
\includegraphics[width=0.45\textwidth]{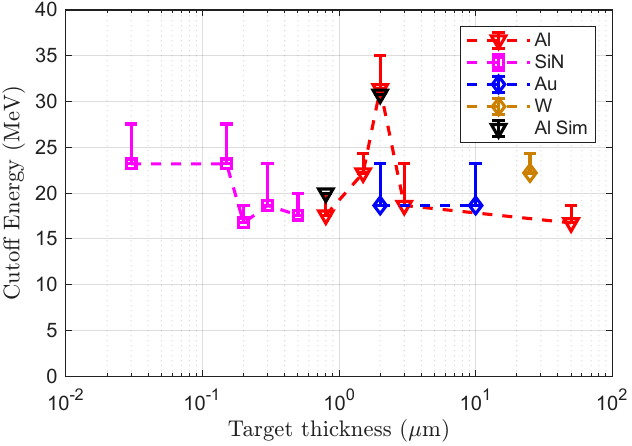}
\caption{Proton cutoff energy as a function of target thickness (for different target materials, see legend). The data points indicate the last RCF layer with clear proton signal and the upper error bars represent the energy corresponding to the next RCF without proton signals. The two black empty triangles represent the corresponding simulation results.}
\label{thickness}
\end{figure}

Protons emitted from the target rear were recorded using a stack of radiochromic films (RCF) \cite{bolton2014instrumentation} positioned along the target normal [Fig.~\ref{exp_setup}]. These protons originate from hydrogenated surface contaminants (e.g., water vapor) rather than the bulk target material. The maximum proton energies inferred from the RCF measurements are shown in Fig.~\ref{thickness} as a function of target thickness.

We identified two distinct ion acceleration scenarios, consistent with our previous experimental results \cite{antici2009laser, antici2017acceleration}. First, for thick targets with intact solid surfaces (e.g., the $2\,\rm \mu m$ Al case), protons are accelerated via the so-called Target Normal Sheath Acceleration (TNSA) mechanism, that is, by the intense sheath field created by hot electrons expanding into the vacuum, resulting in the highest observed proton cutoff energy of 31.4~MeV. This result is in agreement with the theoretical scaling for TNSA \cite{Lacoste2026}, which yields a 31.6~MeV cutoff for a 140~nm pre-plasma scale length. 

On the other hand, when the laser interacts with a thin target that has been decompressed by the laser prepulse into a large-scale, low-density plasma (e.g., the 30~nm and 150~nm SiN foils), Collisionless Shock Acceleration (CSA) sets in: protons are then picked up and accelerated within the target volume by a collisionless shock wave launched at the target front and propagating through the plasma. This mechanism yielded the second-highest cutoff energy of 23.3~MeV.

\begin{figure}[hbtp]
\centering
\includegraphics[trim={0cm 0cm 0cm 0cm},clip,width=0.45\textwidth]{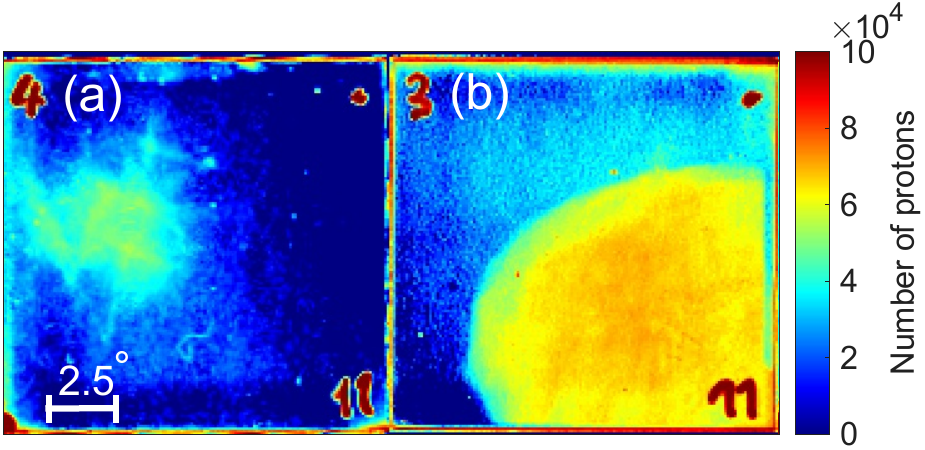}

\caption{Experimental measurements of proton beam profiles, recorded from  (a) a 30 nm SiN target and (b) a 2 $\mu$m Al target using RCF films \cite{bolton2014instrumentation}. The films contain a polymer which darkens when ionizing radiation deposits its energy within. It thus allows to obtain a continuous profile of the incoming radiation on the film, as shown. For both of these two shots we show the 11th layer in the RCF stack, which corresponds to 23.2 MeV of proton energy. 
}
\label{beam_profile}
\end{figure}

The transition between the two acceleration regimes for nm- and $\rm \mu m$-thick targets is further evidenced by the starkly different spatial distributions of the proton beams. Figure~\ref{beam_profile} shows two illustrative proton beam profiles: one from a 30~nm thick SiN foil [Fig.~\ref{beam_profile}(a)] and one from a $2\,\rm \mu m$ thick Al foil [Fig.~\ref{beam_profile}(b)]. The significantly higher collimation observed in the thin-target case aligns with our previous findings \cite{antici2009laser, antici2017acceleration} and provides additional evidence that proton acceleration is dominated by CSA for thin targets and TNSA for thick targets.


\begin{figure}[hbtp]
\centering
\includegraphics[width=0.45\textwidth]
{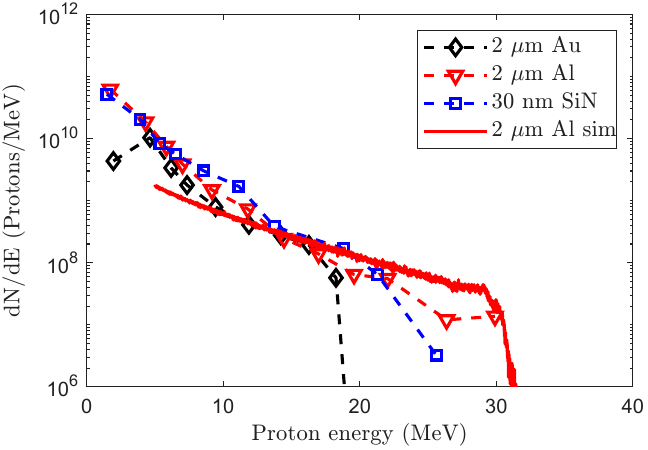}
\caption{A subset of experimental measurements of proton energy spectra analyzed from RCF data, as well as 2D PIC simulation results 
(see text for details).
}
\label{proton_spec}
\end{figure}

Figure~\ref{proton_spec} displays the proton energy spectra measured using various target configurations. The proton spectra recorded from $\rm \mu m$ thick foils in the TNSA regime are in remarkable agreement with those obtained using the 1~PW laser beam at the Apollon facility \cite{lelievre2024comprehensive,Burdonov2021-jc,Yao2024-ti}, highlighting the consistency and robustness of the proton beams  produced across high-power laser systems with similar parameters. Table~\ref{tab:Result_proton} summarizes the laser-to-proton conversion efficiencies for different targets, indicating that the efficiency remains comparable between the CSA and TNSA regimes.


\subsubsection{Particle-in-cell simulations.}

Particle-in-cell (PIC) simulations are performed in two-dimensional (2D) geometry, using the fully relativistic \textsc{SMILEI} code \cite{derouillat2018smilei}. The simulation domain has dimensions $L_x = L_y = 160$ $\mu$m, with spatial resolution $d_x = d_y = 31.25$ nm, thus having the number of cells $N_x = N_y = 5120$. We put 100 particles per cell.
It is known that 2D simulations tend to overestimate the laser-target coupling and the resulting proton acceleration compared to 3D \cite{Brantov2011}, but 
full 3D simulations are very computationally expensive. Hence, to manage simulations in 2D while having realistic results, we anchored our 2D numerical modeling to the experiment by using the cutoff energy measured from the $2\,\mu\mathrm{m}$ and $0.8\,\mu\mathrm{m}$ Al targets (see Fig.~\ref{thickness}), as well as the electron spectra (see Fig.~\ref{e-spec}).
The anchoring procedure is done as follows:
the laser energy input in the simulation and the scale length of the pre-plasma at the target front (assuming an exponential shape, induced by pre-pulse irradiation of the target) are fixed by anchoring the simulation results to the measurements of the proton cutoff energy and of the electron spectra. 
Using the $2\,\mu\mathrm{m}$ Al target, we scanned the laser energy input into the 2D PIC simulation, as well as the scale length of the preplasma, so that we arrived at a measured cutoff energy about 34 MeV (with 6 J of laser energy in the simulation), i.e., similar to that recorded in the experiment, with as well a similar hot electron temperature (with an exponential scale-length of $24\,\mu\mathrm{m}$ for the pre-plasma).  



\begin{table}[b]
\begin{ruledtabular}
\begin{tabular}{cc}
Target & laser-to-proton conversion efficiency\\
\hline
\hline
        30 nm SiN 
        & 0.55\%
        \\
        \hline
        500 $\mu$m SiN 
        & 0.58\%
        \\
        \hline
        1.5 $\mu$m Al 
        & 0.63\%
        \\
        \hline
        2 $\mu$m Au 
        & 0.52\%
        \\
        \hline
        10 $\mu$m Au 
        & 0.57\%
        \\
        \hline
        25 $\mu$m W 
        & 0.42\%
        \\
\end{tabular}

\end{ruledtabular}
 \caption{Summary of the measured laser-to-proton energy conversion efficiency for various targets, evaluated by integrating (over $0\le E \le E_{cutoff} 
    $
    ) the exponential fit of each recorded proton spectrum.}
    \label{tab:Result_proton}
\end{table}







\subsection{X-ray measurements}\label{Sec_Xray}

\subsubsection{X-ray energy and spectra modeling.}

\begin{figure}[hbtp]
\centering
\includegraphics[ width=0.45\textwidth]{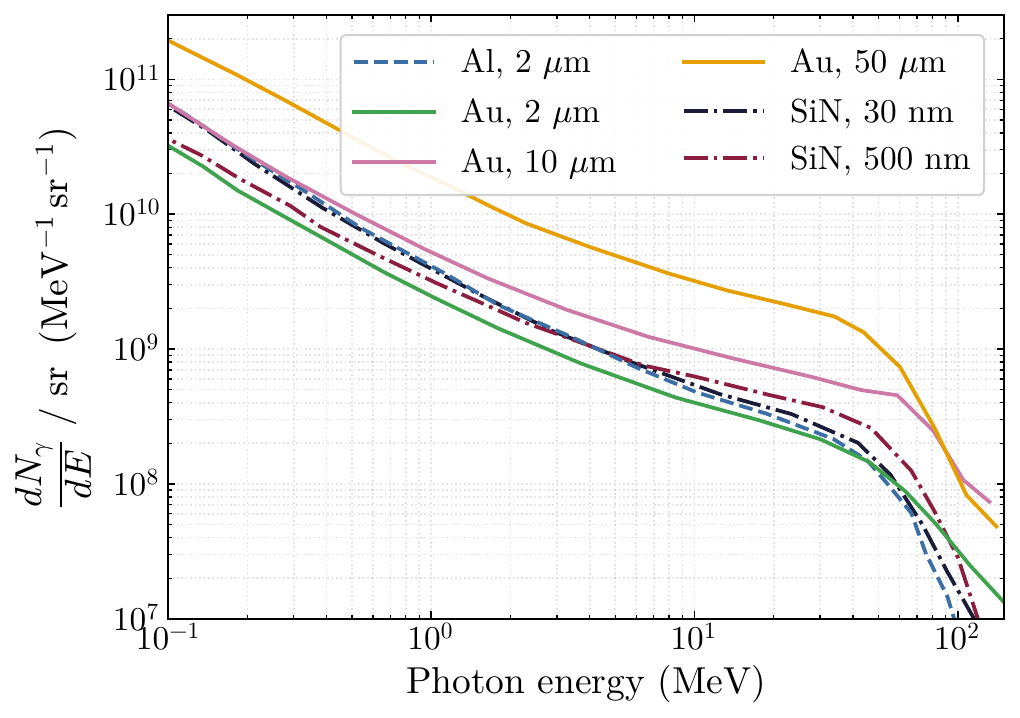}
\caption{Unfolded photon energy spectra for gold, silicon nitride, and aluminium targets of different thicknesses. Typical uncertainties on photon number inside each bin are about 50\% but have been omitted for clarity.}
\label{fig:X_ray_spectra}
\end{figure}

The energy spectra of the X-ray photons generated during the laser-target interaction  were recorded at 18 degrees from the target normal (see Fig.~\ref{exp_setup}). The measurements of the X-rays were performed using stacked scintillators \cite{Fauvel_RSI_25}, the emission of which was recorded by a CCD camera. 
The X-ray spectrometer stack consisted  of 2 layers of plastic scintillators, 10 layers of YAG:Ce and 5 layers of CsI:Tl; as described in \cite{Fauvel_RSI_25}. The detector  covered an opening solid angle of 2.1 $\times 10^{-3}$~sr. The spectrometer enabled detection of photons from 100 keV to 200 MeV. Using a stochastic unfolding algorithm available at \cite{ Fauvel2026-cm}, we could retrieve the photon energy distributions shown in Fig. \ref{fig:X_ray_spectra}.

The obtained spectra display typical features of Bremsstrah\-lung radiation, following a power-law distribution of power -1 with an exponential-like cut-off, as described by Kramer's law \cite{Kramers01111923,hannasch_compact_2021}, around 100 MeV. It is important to note that the unfolding process to obtain the absolute spectra from the raw data tends to slightly overestimate the photon yield where the spectra sharply decline, typically by a factor of three. 

It is interesting to note that 
the electron spectrum shown in Fig.~\ref{e-spec} exhibits a sharper drop at high energy 
for thick targets than for thin targets. 
This leads us to postulate that, for the thick targets, the high energy electrons lost proportionally more of their energy through Bremsstrahlung to generate X-rays
.

\begin{figure}[hbtp]
\centering
\includegraphics[trim={0cm 0cm 0cm 0cm},clip,width=0.45\textwidth]{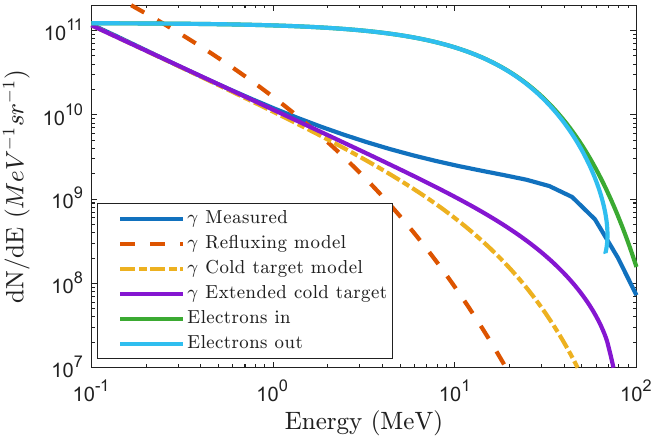}
\caption{Bremsstrahlung calculations for various models of electrons passing through the target (see text for details). The blue line corresponds to the measured X-ray spectrum generated from targets of 25~$\mu$m W. We can observe that the "extended cold target" model (in violet) is the closest to the actual data (in dark blue). 
}
\label{Bremsstrahlung}
\end{figure}

We tested three models of Bremsstrahlung emission within the target, all of which use an electron spectrum with a temperature of 15~MeV in a 25~$\mu$m W target. These are:
i) a "refluxing model": it comprises a full electron refluxing model that takes into account the expansion of the target using the methods described in \cite{Antici2013-ze}, 
ii) a "cold target model": it models only a single pass of the electrons in the target being considered as cold matter.
iii) an "extended cold target": it takes into account a partial refluxing of the electrons as they pass in the target being treated as cold matter. The latter model considers that the refluxing increases with the energy of the electrons. The number of refluxes is parametrized as $\exp(E/E_{\mathrm{reflux}})-1$, where $E_{\mathrm{reflux}}=25 MeV$. The logic behind such dependence is that higher energy electrons  have a longer  effective mean free path as they laterally reflux within the target \cite{Green2018-bo,CompantLaFontaine2013,Iwata2021-nc}. Hence, they effectively crisscross longer through the target and encounter more matter on the way.
The value of $E_{\mathrm{reflex}}$ was chosen to best match the shape of the x-ray spectrum and to recreate the sharp drop that we see in the electron spectra, in Fig.~\ref{e-spec}, at higher energies for the case of the thick targets of a few 10~$\mu m$, at roughly the energy of $E_{\mathrm{reflex}}$.
For these calculations, the Bremsstrahlung cross section was taken by interpolating the measured results given in Ref.~\cite{Seltzer1985}
The result of these models is shown in Fig.~\ref{Bremsstrahlung} along with the experimental X-ray measured spectrum. Obviously, the most accurate reconstruction of the spectra is given by model (iii), for which we also plot the electron spectrum after passing the target. An other point that supports the model is that, with it, the output electron spectrum exhibits a sharper slope at higher energies, similar to the one recorded in the experimental spectra shown  in Fig.~\ref{e-spec}(a).

We further observe in the spectra shown in Fig. \ref{fig:X_ray_spectra} that the cutoff photon energy reaches a maximum for targets around $2$-$3$ $\mu$m gold, consistently with the fact that these exhibit the highest electron temperature. We  note that the photon yield shows much more  variations  across the different target materials than the maximum photon energy. As expected from the $Z^2$ scaling of Bremsstrahlung emission, the high-$Z$ gold targets ($Z = 79$) generate a substantially larger number of photons than aluminium or silicon nitride targets. 

\begin{figure}[hbtp]
\centering
\includegraphics[width=0.4\textwidth]{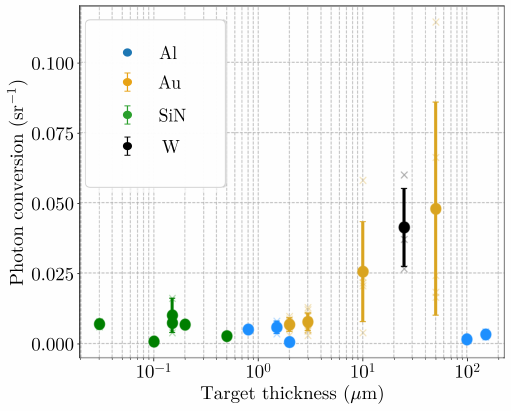}
\caption{Laser-to-photon conversion efficiency per sr for different materials as a function of target thickness, as measured by the detector positioned at $18^{\circ}$ from the laser axis. The vertical error bars correspond to the range of measurements over various shots performed with the same target.}
\label{fig:X_ray_conversion}
\end{figure}

This enhanced photon production is also evident in the laser-to-photon conversion efficiency measured by the detector, which is shown in Fig. \ref{fig:X_ray_conversion}. Gold displays a noticeably higher conversion ratio than aluminium, particularly at large target thicknesses. 

For target thicknesses below approximately $10 \mu$m, the conversion efficiencies remain within a comparable range for all materials. Above this thickness, gold conversion efficiency increases, reaching close to an order of magnitude higher than aluminium or silicon nitride, though the limited shot statistics (N=1–5 per bin) mean this trend is not resolved at the 2$\sigma$ level.


\subsubsection{X-ray imaging.}
\begin{figure}[hbtp]
\centering
\includegraphics[width=0.47\textwidth]{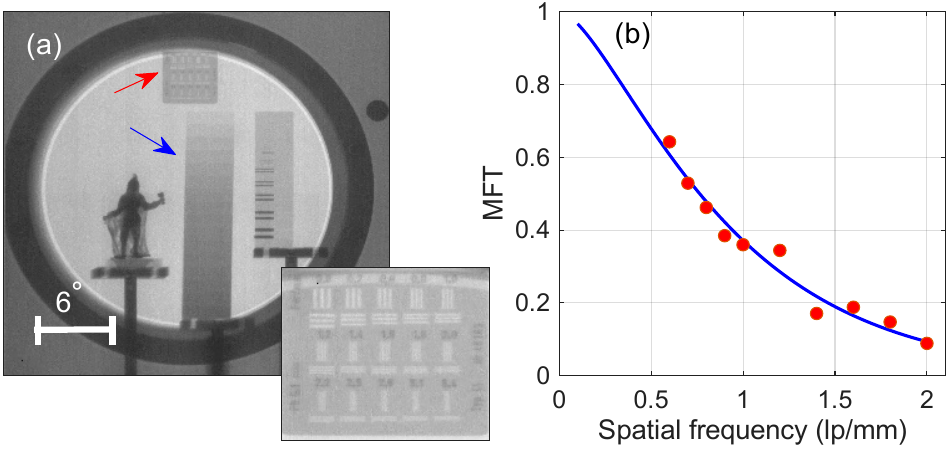}
\caption{(a) X-ray panel image recorded when shooting  a 25 \(\mu m\) W target, with various objects placed in front of the panel, including a line gauge IQI indicated by the red arrow and inlarged in the inset at the bottom right, and an aluminum step block indicated by the blue arrow. (b) Plot of the Modulation Transfer Function (MTF) obtained from the line gauge IQI shown in the red box of panel (a). From the Fourier transform of this plot of the MTF (see text), we can obtain an effective X-ray source size, which is plotted in Fig.~\ref{source size}.}
\label{MTF}
\end{figure}

To assess the imaging capability of the bright X-ray source, the X-ray burst was used to image multiple radiography objects, including a line pair gauge Image Quality Indicator (IQI). The imaging was performed using a PerkinElmer 1621 X-ray panel amorphous silicon panel as shown in Fig.~\ref{MTF} (a).

\begin{figure}[hbtp]
\centering
\includegraphics[ width=0.45\textwidth]{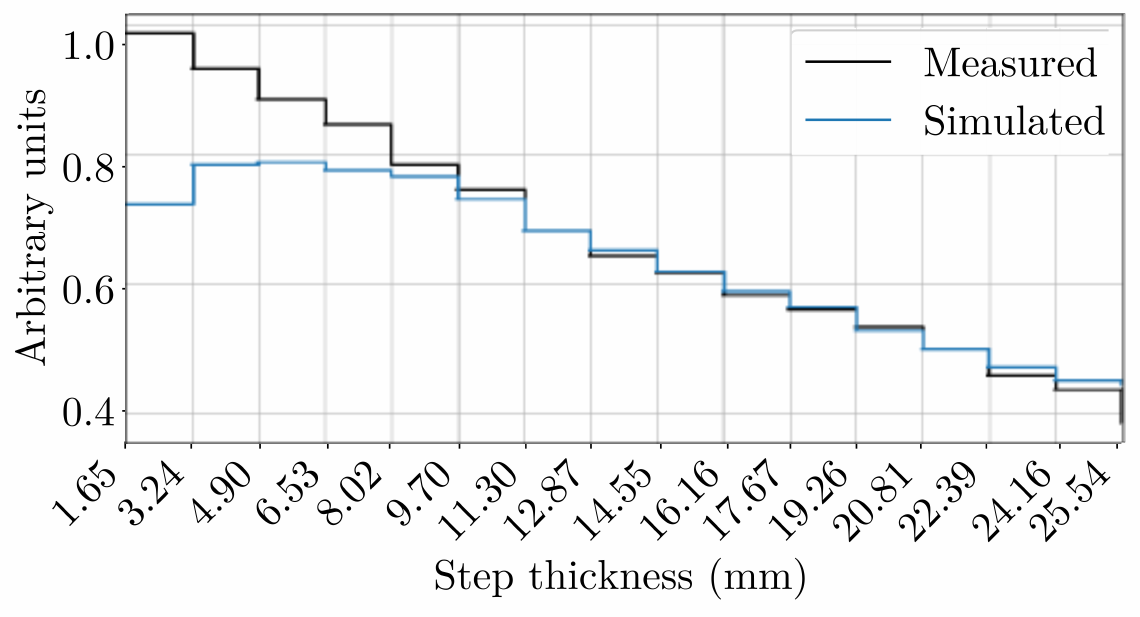}
\caption{Comparison between the  measured (in black) and simulated (in blue) X-ray transmission through the Al wedge shown at the center of the image shown in Fig.~\ref{MTF}. The simulation uses the photon spectrum unfolded for the corresponding target (25 \(\mu m\) W), as shown in Fig. \ref{fig:X_ray_spectra}. Low energy (i.e. below 100 keV) photons are not included in the simulation since they cannot be measured by the  spectrometer, thus explaining the observed discrepancy for the thin steps of the wedge.} 
\label{fig:Wedge_comparison}
\end{figure}

Various objects were radiographed, including an Al wedge (seen at the center of Fig.~\ref{MTF} (a)). The latter allows us to assess the fidelity of the  photon spectra reconstruction. Indeed, we can use it to compare the measured transmission through the wedge to the  simulated one, using the  photon spectra unfolded from the spectrometer (as shown in Fig. \ref{fig:X_ray_spectra}). The comparison is shown  in Fig. \ref{fig:Wedge_comparison}. The Monte-Carlo calculation of the X-ray transmission through the wedge is performed using the FLUKA code\cite{Bohlen2014-wb}.  The simulation setup reproduces the experimental one, including the 1 cm aluminium flange, the wedge step thicknesses, and the various covers in front of the PerkinElmer  panel, i.e. the 500 $\mu m $ Al cover, the 500 $\mu m $ of Cu converter and the 450 $\mu m$ thick scintillators of Gadolinium oxysulfide ($Gd_2O_2S$).

For large step thicknesses, the agreement between the simulation and the data is excellent. A discrepancy appears for the thinnest steps and is readily explained: the detector is not sensitive to photons below 100 keV, so the incident spectrum used for the simulation lacks this low-energy component. In the experiment, these low-energy photons can traverse the thinnest steps and increase the measured signal. As the step thickness grows, they are absorbed upstream and the discrepancy vanishes.

\subsubsection{X-ray source size.}
To measure the  size of the X-ray source, we use the IQI. From it, we can calculate a modulation transfer function (MTF). This is done  for each  shot. 
The MTF results were fitted to the Bennet function, following the approach described in Ref.~\cite{Ekdahl2011}. The analysis was performed using a one-dimensional MTF. Under the assumption of a spatially symmetric X-ray source in the plane of the target, the corresponding optical transfer function (OTF \cite{Williams2002}) is real and symmetric, and its magnitude is therefore equivalent to the MTF used in the analysis. This assumption is consistent with our experimental results, as the MTFs obtained from objects oriented along the horizontal and vertical directions were found to be similar. Consequently, the source PSF can be related to the corresponding OTF through inverse Fourier transformation under this symmetry assumption.
To account for the contribution of the non-uniform X-ray background, a region near the object used for the MTF analysis was selected, and a polynomial function was fitted to the intensity profile in this region. The fitted background function was subsequently subtracted from the object profile prior to the MTF calculation. This procedure reduces the influence of spatial variations in the background originating from the X-ray source.

A MTF, and a fit thereof, is shown in Fig.~\ref{MTF} (b) for a shot on a 25 \(\mu m\) W target. 
Due to the penetration of the high energy X-ray through the thin IQI, the maximum achievable contrast is reduced. To account for this effect, an additional normalization constant was included in the MTF fit.
A Fourier transform was then performed on the MTFs to establish the point spread function (PSF) for each target material of various thicknesses.

\begin{figure}[hbtp]
\centering
\includegraphics[width=0.45\textwidth]{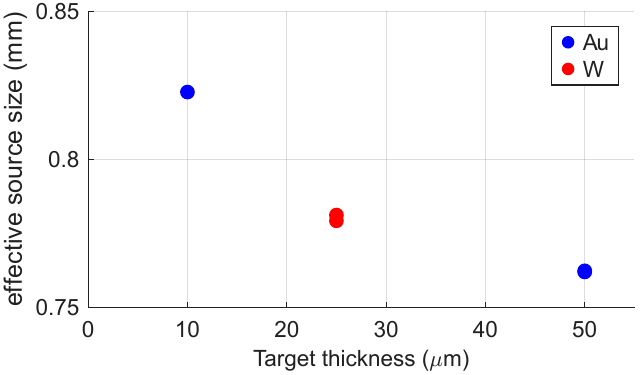}
\caption{Effective X-ray source size measured as a function of target thickness for different materials (see legend). Each data point represents the mean value obtained from multiple measurements performed on  similar targets.
}
\label{source size}
\end{figure}

The effective X-ray source size was finally calculated by taking the FWHM of the PSF for each shot. 
The measured quantity represents the effective spatial resolution (or effective source size) of the complete imaging system, including contributions from both the X-ray source and detector. Consequently, the measured source size should be regarded as an upper limit on the actual X-ray source size.
The source size of the X-ray emission for various targets is shown in Fig.~\ref{source size}. The measured source size varies from 0.76 to 0.82 mm, with thicker foils producing smaller sources and better images. These are somewhat larger, but of the same order of magnitude, than what was recorded in previous experiments performed at various laser facilities \cite{Courtois2013-af,Brenner2016-vi,Palaniyappan2018-kx,Edwards2002-oi,Hollinger:25}. 
These sizes are also much larger than the laser focal spot due to several effects. First, the electrons recirculate within the target and laterally  expand far from the focal spot \cite{Green2018-bo,CompantLaFontaine2013,Iwata2021-nc}, inducing X-ray emission over a large area. Second, because of the limited contrast of the ELI-NP laser system, the pedestal before the main pulse has sufficient energy to vaporize the thin foils and the plasma has begun to expand. When the main pulse arrives approximately 100 ps later (see Fig.~\ref{laser}(b)) and heats the target plasma, the X-ray emission area is significantly larger than the laser focal spot size. We must also consider that the temporal gating of the flat panel detector is essentially infinite, 1 second, on the time scale of the laser-plasma interaction. Thus,  X-rays are continuously produced as the plasma continues to expand and cool, and  are  detected and incorporated into this spot size measurement.

\subsubsection{X-ray angular distribution.}
\begin{figure}[hbtp]
\centering
\includegraphics[trim={0cm 5cm 0cm 5cm},clip,width=0.45\textwidth]{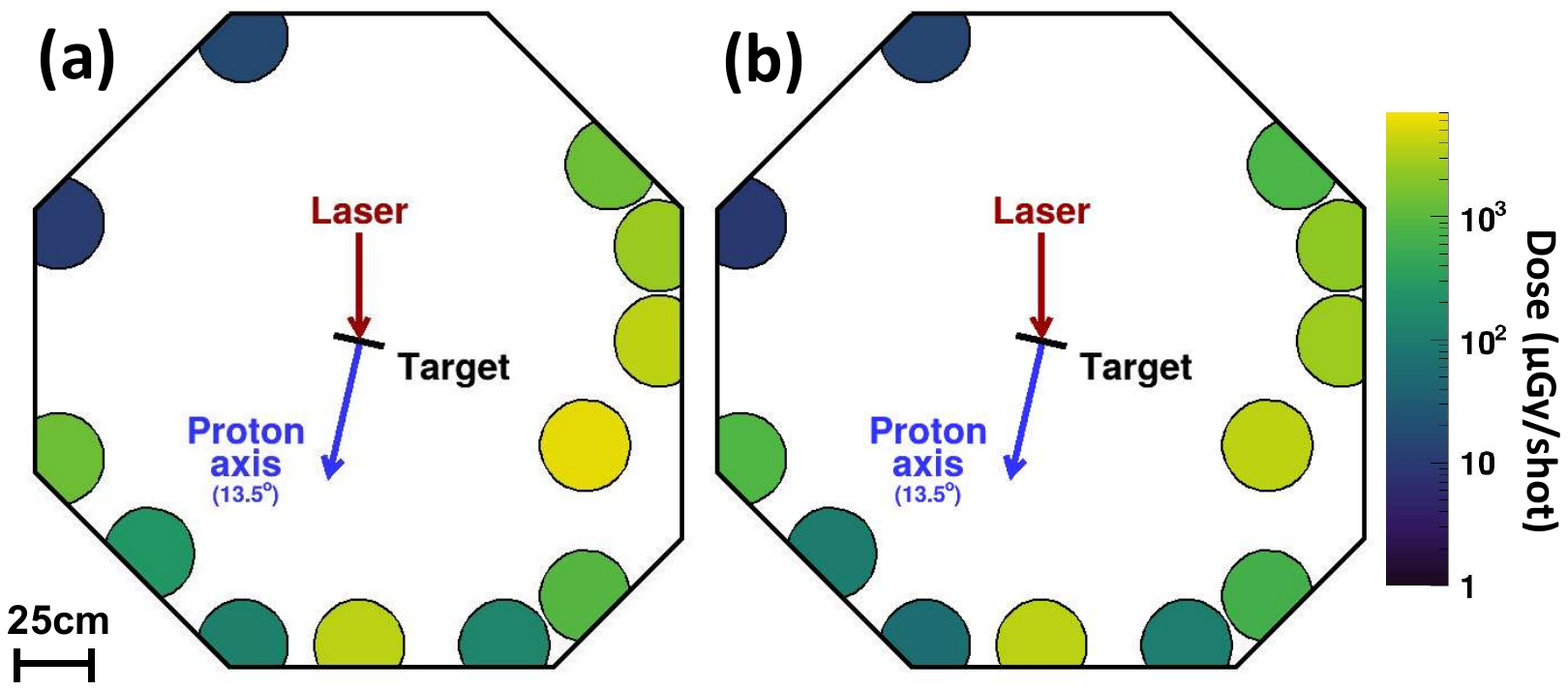}
\caption{X-ray spatial dose distribution measured inside the interaction chamber, using RPL dosimeters and for shots on (a) $25\,\mu\mathrm{m}\, \mathrm{W}$ and (b) $30\,\mathrm{nm}\,\mathrm{SiN}$.
Each circle represents the position of a dosimeter, and its color corresponds to the measured dose.
}
\label{dosi}
\end{figure}

As shown in Fig.~\ref{dosi}, the angular distribution of the X-rays has been mapped using radiophotoluminescence (RPL) dosimeters \cite{Ferrari2024-wm} positioned inside the target chamber. By comparing the results obtained from $25\,\mu\mathrm{m}\, \mathrm{W}$ and $30\,\mathrm{nm}\,\mathrm{SiN}$, we can see that no clear angular distribution difference between them is observed
.
The angular distribution also shows two preferred directions for emission, along the laser axis and along  the target surface. It has been recently shown in simulations that the emission along the target surface is due to the interference pattern in the electromagnetic field formed by the incident and reflected laser pulse. The electromagnetic field accelerates electrons, while temporarily directing their momentum along the target surface. Consequently, they emit a collimated $\gamma$-photon beam in the same direction \cite{Matys2025}.

\subsection{Neutron measurements}\label{Sec_Neutrons}
\subsubsection{Neutron energy spectra and modeling.}
Neutrons were generated via proton interactions, in a pitcher-catcher configuration, with a 4~mm thick and 25.4~mm diameter LiF converter, positioned 12 mm downstream of the laser-irradiated target. 

\begin{figure}[hbtp]
    \centering
    \includegraphics[width=1\linewidth]{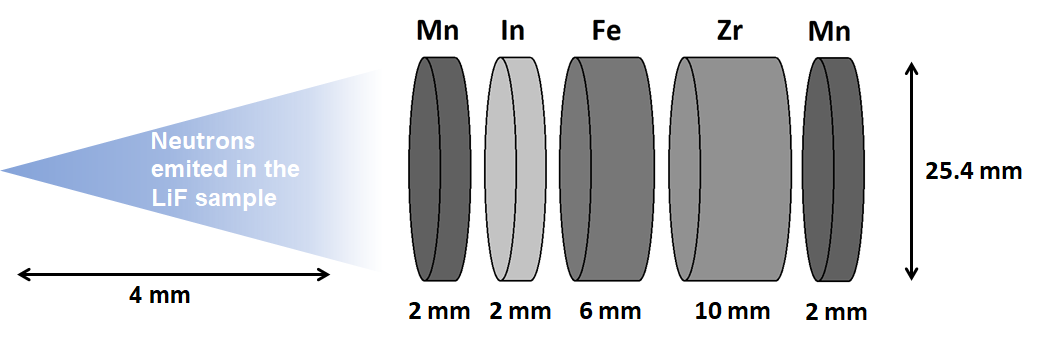}
    \caption{Design of the SPAC activation diagnostic used during the experimental campaign. The elements are chosen to have complementary energy range of activation induced by the incoming neutrons, as well as to minimize activation of each sample by the radiation from the other samples \cite{Lelievre_thesis}.}
    \label{fig:SPAC_design}
\end{figure}

Neutron production in the LiF converter arises from two nuclear reactions : $^7\rm Li(p,n)^7Be$ and $^{19}\rm F(p,n)^{19}Ne$. The selected converter thickness ensures that all protons with energies below 29.2 MeV interact within the material. Since no higher-energy protons were measured, all protons are expected to interact within the LiF converter.
\\
The neutron emissions were characterized using activation measurements. A neutron activation spectrometer (SPAC), consisting of a stack of metallic foils with accurately known isotopic compositions arranged in order of increasing reaction threshold energy, was used. Positioned directly behind the LiF converter, the spectrometer intercepted the largest possible fraction of the emitted neutrons, thereby maximizing the activation yield. A schematic representation of the diagnostic is shown in Fig.~\ref{fig:SPAC_design}, and the main nuclear reactions associated with each activation foil are summarized in Table~\ref{tab:Result_neutron}.

Since these neutrons are mainly produced through the $^7\rm Li(p,n)^7Be$ reaction~\cite{lelievre2024comprehensive}, the total number of emitted neutrons can be estimated by measuring the quantity of $^7\rm Be$ nuclei produced in the converter.

The activation measurements were performed using semiconductor gamma-ray spectrometers: one Broad Energy Germanium (BEGe) detector and two High Purity Germanium (HPGe) detectors. Their energy and efficiency calibration was carried out using calibration point sources of $^{54}Mn$, $^{57}\rm Co$, $^{60}\rm Co$, $^{133}\rm Ba$ and $^{137}\rm Cs$. The absolute efficiencies were then extrapolated to the geometries of the LiF converter and the foils using Geant4 simulations~\cite{AGOSTINELLI2003250}.
The acquired gamma-ray spectra were analyzed using the ROOT framework \cite{ROOT_NIMA_1997} and the activities were calculated using the following equation:
\begin{equation}
    A_{\rm meas}=\frac{S_{\rm net}}{I_\gamma \times\epsilon}\times e^{\lambda t_{\rm decay}}\times\frac{\lambda}{\left(1-e^{-\lambda t_{\rm meas}}\right)} 
\end{equation}
Considering $t_{\rm decay}$ as the time passed between the last laser shot and the start of the gamma acquisition, and $t_{\rm meas}$ the measurement duration. $S_{\rm net}$ denotes the net peak area, obtained after subtraction of the background contribution from the gross peak area. $I_\gamma$ is the emission intensity of the $\gamma$ ray under consideration, and $\epsilon$ is the detector efficiency at the corresponding $\gamma$-ray energy for the experimental configuration used. \textit{$\lambda$} is the decay constant ($s^{-1}$) of the considered emission.

\subsubsection{Monte Carlo simulations.}
The neutron production was first simulated using Geant4 with the ENDF/B-VIII.0 cross-section library~\cite{BROWN20181}, under the same configuration as the experiment. Specifically, the simulations considered the measured proton spectrum obtained with the 2 $\mu$m Au target (shown in Fig.~\ref{proton_spec}), which is representative of the targets used during neutron production. These protons were injected into a 4~mm thick LiF converter with a half angle of 21°, thus reproducing the beam divergence of a TNSA proton beam~\cite{mancic}.

\begin{figure}[ht]
    \centering
    \includegraphics[width=1\linewidth]{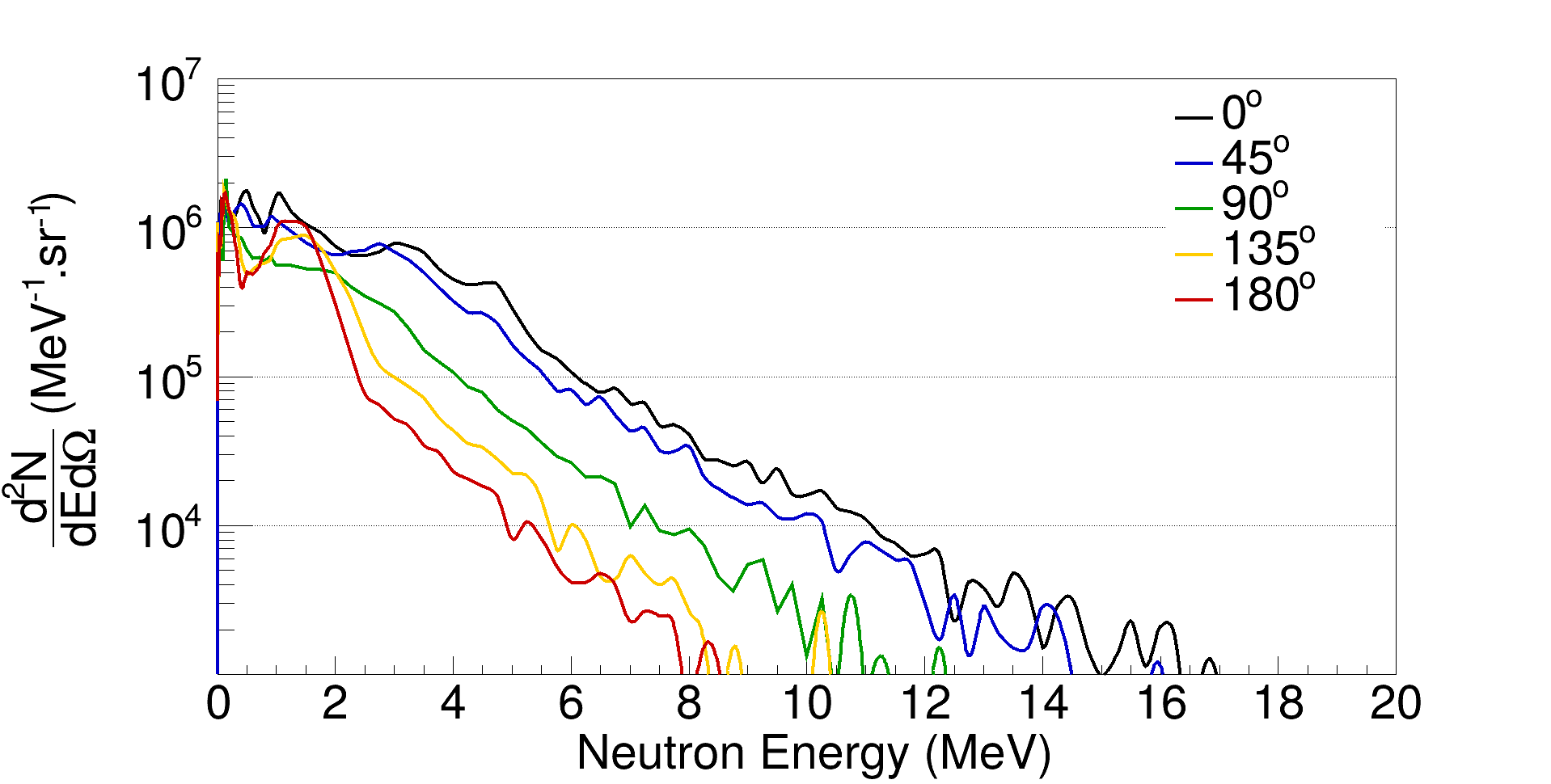}
    \caption{Simulated neutron energy spectra at different angles, produced by the interaction of protons generated from a 2 $\mu$m thick Au target with a 4-mm-thick LiF converter.}
    \label{fig:Neutron_spectra}
\end{figure}

The simulated neutron energy spectra at different angles compared to the target normal axis are shown in Fig.~\ref{fig:Neutron_spectra}. Although the neutron energy spectrum appears identical between 0° and 45°, these simulation results reveal a clear anisotropy at higher emission angles, with more neutrons of higher energy emitted forward than backward.

\begin{table}[b]
\begin{ruledtabular}

\begin{tabular}{ccccc}
       
\hline
\textbf{Materials}
& \makecell{$\mathrm{A_{meas}}$ \\ $/\,\mathrm{shot}$ (Bq)}
& \makecell{$\sigma_{\mathrm{meas}}$ \\ (\%)}
& \makecell{$\mathrm{A_{sim}}$ \\ (Bq)}
& \makecell{Rel. diff. \\ (\%)} \\
\hline
\hline
        
        \textbf{LiF}\\
$^{7}\mathrm{Li}(p,n)^{7}\mathrm{Be}$& 8.47
        & 8.21
        &8.76& -3.52\\
        \hline
        \textbf{Manganese}\\
$^{55}\mathrm{Mn}(n,g)^{56}\mathrm{Mn}$& 0.18
        & 33.61
        & 0.167
        & 6.65\\
        \hline
        \textbf{Indium}\\
$^{115}\mathrm{In}(n,g)^{116m}\mathrm{In}$& 4.83
        & 13.58
        & 4.02& 16.86\\
        \hline
        \textbf{Indium}\\
$^{115}\mathrm{In}(n,n')^{115m}\mathrm{In}$& 2.14
        & 13.00
        & 2.21& -3.52\\
        \hline
        \textbf{Iron}\\
$^{56}\mathrm{Fe}(n,p)^{56}\mathrm{Mn}$& 0.15
        & 20.29
        & 0.136& 12.45\\
        \hline
        \textbf{Zirconium}\\
$^{90}\mathrm{Zr}(n,2n)^{89}\mathrm{Zr}$& 0.07
        & 19.99
        & 9.28E-02& -27.57\\
        \hline
        \textbf{Manganese}\\ 
$^{55}\mathrm{Mn}(n,g)^{56}\mathrm{Mn}$& $<$ \textit{DL}*& $\varnothing$
        & 0.0131
        & $\varnothing$
            \\
    \end{tabular}
    \end{ruledtabular}
    \caption{Summary of the measured and simulated activities of the LiF converter and the samples within the SPAC spectrometer following neutron irradiation (see text for details). 
     *\textit{DL stands for Detection Limit.}} 
    \label{tab:Result_neutron}
\end{table}

\subsubsection{Activation data.}
The LiF converter was used during a series of 5 shots, yielding a measured $^7\rm Be$ activity of 8.47 ± 0.69 Bq/shot. Given that 96\% of the neutrons are produced via the $^7\rm Li(p,n)^7Be$ reaction, as previously determined, the total neutron production is estimated at $(5.85\,\pm\,0.48)\times 10^{7}$ neutrons/shot. This value is 3.5~\% lower than that predicted by the Geant4 simulations. It is consistent with the results obtained under the same configuration using the secondary laser beam at the Apollon facility~\cite{lelievre2024comprehensive}.

The SPAC was used during the same series of 5 shots as the LiF converter and the activity per shot of each foil, corrected from the radioactive decay between successive shots, are shown in Table~\ref{tab:Result_neutron}. For the measurement, uncertainties associated with the peak area determination and the efficiency calculations were taken into account.

The neutron spectra presented  in Fig.~\ref{fig:Neutron_spectra} 
were  used as input for activation simulations of the foils, performed using the MCNP6 Monte Carlo transport code \cite{Rising2025-zu} with the IRDFF-II cross-section library \cite{Trkov2020-wa}. This library was selected based on previous activation studies, which demonstrated that it is the most suitable library for modeling the activation processes in the foils used~\cite{Lelievre_thesis}.


Besides the activation induced in the samples by the neutrons, several other activation channels were also taken into account. These include activation induced by: (i) primary photons (see Fig.~\ref{fig:X_ray_spectra}) and electrons (see Fig.~\ref{e-spec}) generated following the laser-irradiation of the primary target, (ii) bremsstrahlung photons produced through the interaction of the primary  electrons with the converter, and (iii)  neutrons produced,  via ($\gamma$,n) reactions, by primary photons in the converter and activation samples themselves. 
The photonuclear reactions we considered 
are summarized in Table~\ref{tab}. Overall, the simulated activation of the samples, taking into account all the considered activation channels is  
summarized in Table~\ref{tab:Result_neutron}.

\begin{table}[hbt]
\centering
\begin{tabular}{c c}

\hline
\hline
\hline
\textbf{Material} & \textbf{Reaction}\\
\hline
\hline
LiF converter & $^9\textrm{Li}(\gamma,n)^8\textrm{Li}$ \\
\hline
LiF converter & $^{19}\textrm{F}(\gamma,n)^{18}\textrm{F}$ \\
\hline
Indium sample & $^{115}\mathrm{In}(\gamma,\gamma')^{115m}\mathrm{In}$ \\
\hline
Iron sample & $^{57}\mathrm{Fe}(\gamma,p)^{56}\mathrm{Mn}$ \\
\hline
Zirconium sample & $^{90}\mathrm{Zr}(\gamma,n)^{89}\mathrm{Zr}$\\
\hline
\end{tabular}
\caption{Principal photonuclear reactions considered.}
\label{tab}
\end{table}

\begin{figure}[ht!]
    \centering
    \includegraphics[width=1\linewidth]{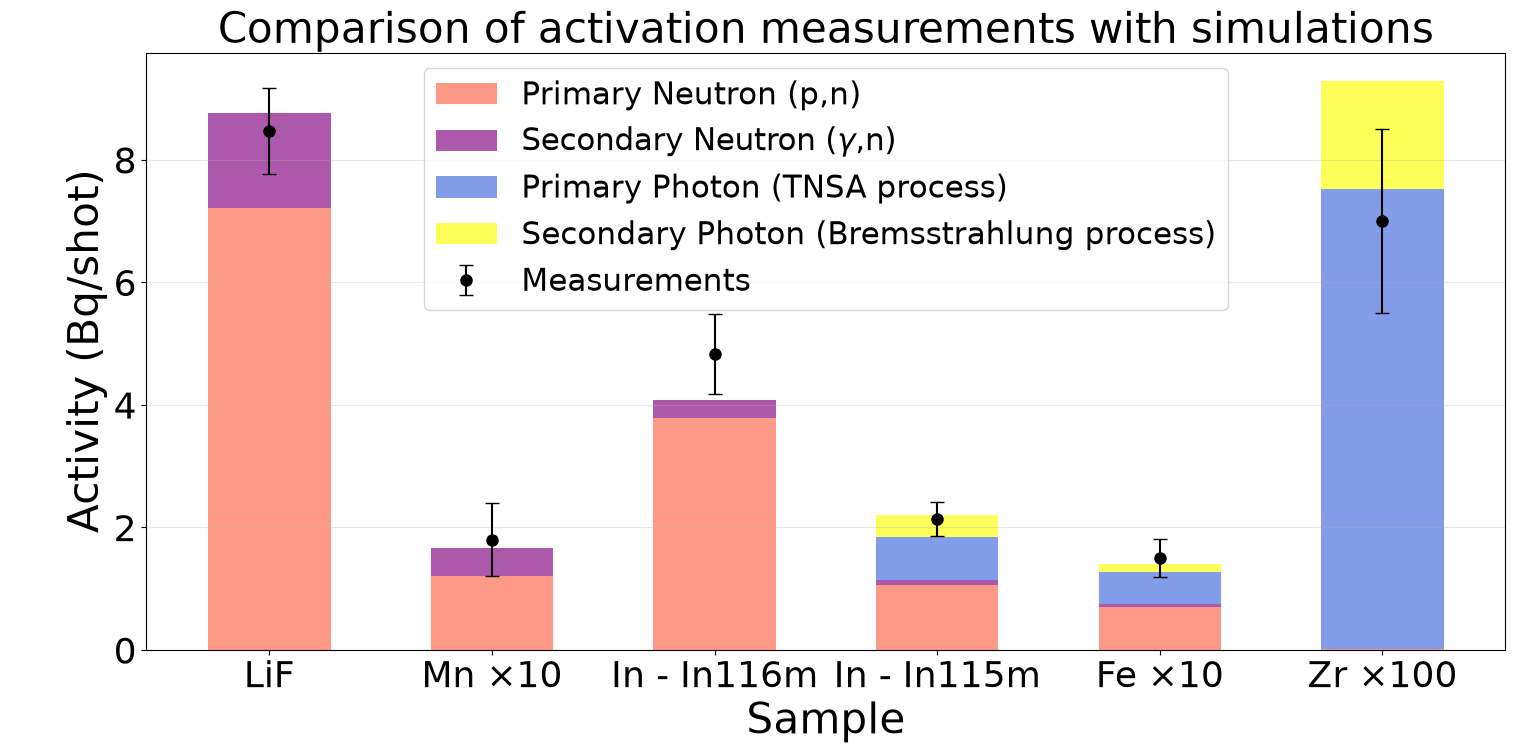}
    \caption{Comparison of measured and simulated activities in the SPAC activation stack, with the details of the different contributions from individual activation channels, as indicated in the legend.}
    \label{fig:activation_result}
\end{figure}

Fig.~\ref{fig:activation_result} summarizes the contribution of the different activation channels into the overall activation of the various samples. The good agreement obtained between the calculated and measured activities demonstrates the importance of accounting for all relevant activation mechanisms in the simulations, besides 
the dominant activation induced by the (p,n) reactions. 
Some discrepancies nevertheless remain, particularly for the $^{89}\textrm{Zr}$ and $^{116m}\textrm{In}$ activities, for which differences of approximately 20\% are still observed but remain reasonable. We also observe that the zirconium activity is mainly due to photonuclear reactions, which the simulations correctly accounts for, through the giant dipole resonance (GDR) \cite{Spicer1969}  ($\gamma$,n) reactions induced by high-energy photons.

Overall, considering all these activation sources, the total simulated neutron yield is $5.81\times 10^{7}$ neutrons/shot, which is in good agreement with the above-mentioned direct measurement of the $^7\rm Be$ nuclei produced in the converter. 

Finally, considering the fact that the cross section of the $^{115}\mathrm{In(n,n')}^{115m}\mathrm{In}$ reaction is significant and almost constant between 1 MeV and 10 MeV, the $^\mathrm{115m}\rm In$ activity induced in the indium foil can also be used to estimate the neutron fluence in this energy range using the following equation:
\begin{equation}
    \Phi = \frac{A_{\rm meas}}{\lambda\times n_{\rm shot}} \times \frac{1}{\overline{\sigma} \times t \times n \times \chi} \times\frac{1}{\Omega}
\end{equation}
where $\Phi$ is the neutron fluence (neutrons/sr/shot), $A_{\rm meas}$ the measured activity (Bq), \textit{$\lambda$} the decay constant ($s^{-1}$), $n_{\rm shot}$ the number of shots, $\overline{\sigma}$ the average cross section ($m^2$) over the 1-10 MeV energy range, \textit{t} the foil thickness (m), \textit{n} the number density ($m^{-3}$), $\chi$ the isotope abundance, and $\Omega$ the solid angle covered by the foil (sr).

From the measured $^{115m}$In activity, a neutron fluence of $(6.38\,\pm\,1.53)\times 10^{6}$ neutrons/sr/shot was inferred in the 1–10 MeV energy range. This value is slightly higher than the simulated fluence of $6.60\times 10^{6}$ neutrons/sr/shot predicted for the same energy range. The uncertainty on the fluence derived from the measurement is calculated as the root mean square of the uncertainty associated to the activity determination, which is estimated to be about 13\%, and the standard deviation of the average cross section $\overline{\sigma}$ which is 20.4\%. 






\subsection{Ongoing prospects for material interrogation}\label{Sec_interrogation}

Recent work have shown the adequacy of MeV-photons, laser-driven source to perform high-resolution tomography of dense objects \cite{Hollinger:25}. Such photon probing has the capability to resolve of the order of a few percent density difference in embedded materials \cite{Kitazawa2005-to}.  A peculiar case is the one of nuclear waste, where metals, organics and liquids can be mixed. Coupling X-ray tomography to measurements of gamma-emitting elements  allows to refine inspection \cite{Bernardi1995-pn}. This would even be further improved by element identification, as uniquely allowed by Neutron Resonance Transmission Analysis (NRTA) setup of heavy elements \cite{Postma2017-yn}. NRTA exploits strong absorption, by specific nuclear resonances, of a  neutron beam traversing an object. Since  resonances are specific to each element, NRTA allows  to uniquely tag elements inside the object. It however requires to use epithermal neutrons (1 eV – 1 keV), since this is where the resonances lead to strong absorption.

In order to verify that the  laser-driven neutron fluxes measured here 
are adequate for probing  nuclear materials using NRTA, we performed simulations of  material interrogation using the PW-scale neutron spectra measured here. This corresponds to the spectra shown in Fig~\ref{fig:Neutron_spectra}, or equivalently to the ones recorded at Apollon \cite{lelievre2024comprehensive}, since the two are very similar.

However, before simulating the material interrogation, we need to simulate the moderation of the neutrons in order to bring the input fast (MeV) neutrons down to the epithermal range. 
The simulated moderator is made of high-density polyethylene (HDPE) and actually includes a cylindrical insert  to hold the LiF converter (having the same characteristics as in the experiment described above) that converts protons to neutrons. The simulation of the moderation is performed using Geant4. Fig.16 shows various moderated neutron spectra, as a function of various length (L) and diameter (D) of the moderator cylinder. The spectra are shown as collected outside the laser-target vacuum chamber, such that we could position there the objects to interrogate. We clearly observe the effect of the moderator in increasing the epithermal neutron fraction, compared to the unmoderated configuration. We note that resonance-like structures at higher energies are also visible, mainly due to scattering and absorption effects in the walls of the aluminum laser-target vacuum chamber. The moderator dimensions were optimized by considering the compromise between epithermal neutron fluence, moderator-induced temporal broadening, and practical geometrical constraints. Increasing the moderator thickness enhances the low-energy neutron fluence, but also increases the neutron residence-time distribution through multiple scattering, thereby degrading the energy resolution of the Time-of-Flight (ToF) detection that can be used in order to retrieve the spectrum of the neutrons \cite{yogo2023laser,lelievre2024comprehensive}. The best compromise was obtained for a moderator diameter of about 10 cm and a length between 3 and 5 cm.
We observe that the epithermal contribution remains small compared to the total neutron yield, i.e. the neutron flux in the epithermal range is reduced by 1-2 orders of magnitude compared to the fast (MeV) neutrons. 

\begin{figure}[hbtp]
    \centering
    \includegraphics[width=0.9\linewidth]{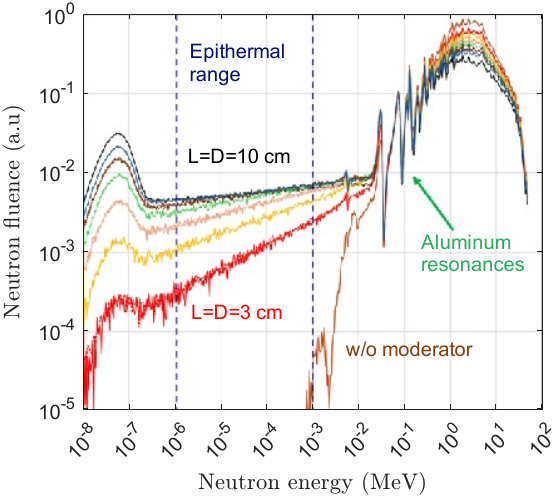}
    \caption{Simulated neutron energy spectra after moderation (see text) and as collected 
    at 1.4 m from the MeV neutron source, for different moderator diameters D and length L
    .}
    \label{fig:NeutronSpectraSim}
\end{figure}

\begin{figure}[hbtp]
    \centering
    \includegraphics[width=0.9\linewidth]{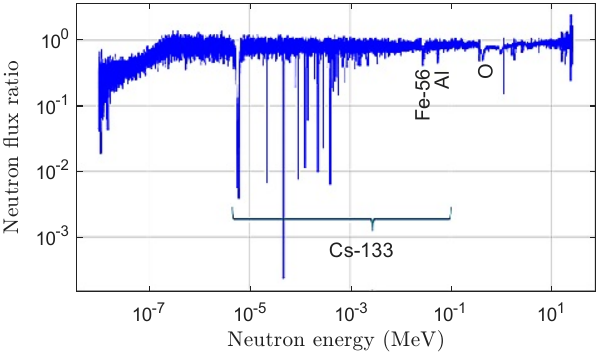}
    \caption{Neutron flux obtained from a Monte-Carlo simulation using PHITS and using the moderated neutron spectrum shown in Fig.~\ref{fig:NeutronSpectraSim}. 
    The neutrons are propagated through a sample consisting of 1 cm concrete, 1 mm Fe, and 1 mm Cs-133.}
    \label{fig:NeutronFluxRatio}
\end{figure}

We then simulated the propagation of the moderated neutrons into various materials. The simulation was performed using the numerical code Particle and Heavy Ion Transport code System (PHITS) \cite{Sato2023}. 
Fig.~\ref{fig:NeutronFluxRatio} shows the result of a simulation 
of NRTA through a sample consisting of 1 cm concrete, 1 mm Fe, and 1 mm Cs-133. Concrete on its own is a heterogeneous material, with elements like Al, Si, H, O, C, Ca and even Fe. For this simulation, we used 3x the neutron spectrum, i.e. it corresponds to accumulating data  over 3 shots. We clearly observe that the particular isotopes contained within the considered object can be identified, owing to their absorption lines in the neutron spectrum. The corresponding characteristic absorption peaks can be found easily, given that  the heights of these peaks depend on the cross-section themselves as well as the mass proportion of the present isotopes. Several aspects are observed: the peaks of concrete are present, the peak of Fe-56 is subtle
, and the peaks of Cs are mainly between the eV - keV, with the peak at 8 eV being the main one. This preliminary calculation illustrates the potential of PW-class laser-driven sources to identify resonance signatures from several isotopes stacked upon one another using only a few shots. However, Fig.~17 represents an idealized energy-domain calculation and does not include the complete experimental ToF response, including the finite detector dimensions and timing response, nor all environmental scattering effects. The resonance structures shown in Fig.~17 should therefore be regarded as a proof-of-principle rather than as the exact spectrum expected experimentally. As demonstrated by other groups, such   transmission measurement can be performed either in a spectrally-resolved mode, using a neutron time-of-flight technique \cite{yogo2023laser}, or in an imaging mode, using a gated imager \cite{Tremsin2014-tk}. The latter has the advantage of providing spatial information but will be restricted, in a single shot, to measure a specific element, by selecting a specific energy in the broad neutron spectrum. Probing the same material using the bright X-rays that we have here characterised will further allow identifying the elements embedded within the probed objects through powerful differential X-ray/neutron absorption \cite{Eberhardt2006-ym}.

An important parameter for NRTA is the energy resolution of the neutron ToF measurement. For a neutron of energy $E_n$, measured over a flight distance $D$ with a time of flight $\tau$, the neutron energy is given by $E_n = \frac{m_n}{2}\left(\frac{D}{\tau}\right)^2$, and the corresponding relative energy resolution can be estimated as $\frac{\Delta E_n}{E_n} = \frac{2\Delta\tau}{D}\sqrt{\frac{2E_n}{m_n}} = 2\frac{\Delta\tau}{\tau}$. Following the procedure used by Yogo et al.\cite{yogo2023laser}, the total temporal spread can be expressed as $\Delta\tau = \sqrt{\Delta\tau_m^2+\Delta\tau_d^2}$, where $\Delta\tau_m$ represents the temporal broadening induced by neutron moderation and $\Delta\tau_d$ accounts for the detector contribution, including its finite dimensions and timing response. In the present configuration, the initial neutron-production timescale is much shorter than the temporal broadening introduced by moderation in the epithermal range and therefore provides a negligible contribution to the overall resolution. The moderator thickness, detector geometry and flight distance must therefore be optimized together. Increasing the moderator thickness increases the epithermal neutron fluence but also increases $\Delta\tau_m$, whereas increasing the flight distance improves the relative ToF resolution at the expense of neutron statistics at the detector. With the geometry considered here, i.e. a flight distance of approximately 1.4 m and an HDPE moderator with $D=10$ cm and $L=3$--$5$ cm, our preliminary Monte Carlo calculations indicate an energy resolution below approximately 5\% for neutron energies above 0.2 eV. This geometry therefore represents a compromise between epithermal neutron fluence, counting statistics and ToF resolution.

\subsection{Concluding remarks} \label{Sec_Discussion}

Our primary goal in developing  simultaneous laser-driven high energy X-ray and neutron bright sources is radiographic inspection of thick components, in particular by exploiting the differential absorption induced on these two types of radiation \cite{Eberhardt2006-ym}. We have shown here that PW-class lasers, which are now commercially available in compact systems, can produce adequate X-ray and neutron sources for dense material probing. Indeed, laser-driven MeV-range  X-rays have already been shown to be adequate for high-resolution tomography of dense objects \cite{Hollinger:25}, as would be needed to e.g. scan a nuclear package. Further, we have here shown that MeV neutrons, once moderated, enable  NRTA analysis of elements embedded deep inside objects, complementing the first demonstration of laser-enabled NRTA using a large-scale kJ laser \cite{yogo2023laser}, and more recent demonstrations enabled by photoneutron generation which utilize the repetition rate of wakefield electron accelerators \cite{Feng2025}. 
We also note that the recent demonstration of resonance spectroscopy for temperature profiling \cite{Lan2024} and spectroscopy performed using laser-driven fast (MeV range) neutrons \cite{MillnCallado2026} opens the door for complementary application of this technique in that energy window.

As already discussed in Refs. \cite{Yogo2021,Wei2022-mb,Armstrong2024,Treffert2024}, a PW-scale laser-driven dual source  provides key advantages over existing (conventional) sources: (i) a compact system, (ii) pulsed radiation (reducing the radiation exposure), and (iii) a dual and simultaneous X-ray/neutron source. The small source size of laser-driven radiation brings the further advantage of reducing the imaging system size requirements. Combining these two advantages can dramatically reduce the cost of MeV X-ray and neutron imaging facilities. Additional benefits include increased imaging system resolution via micro-focus imaging geometry, the potential for neutron time-of-flight imaging and measurements, as well as dynamic imaging which are supported by the fast pulse inherent to a laser source. A good example of the application space for a micro-focus MeV imaging system is the inspection of additively manufactured metal components \cite{Rao2023-nd}. For these components, X-ray computed tomography is need for high resolution metrology. In this respect, neutron imaging with fast pulse structure accelerators \cite{Nelson2018-um} has shown significant potential, but is only available at large accelerator facilities. Therefore, a laser-driven fast and pulsed source would also enable this type of diagnostic capability at smaller facilities.

\section*{ACKNOWLEDGMENTS}
\begin{acknowledgments}
This work was supported by funding from the European Research Council (ERC) under the European Unions Horizon 2020 research and innovation program (Grant Agreement No. 787539, Project GENESIS), by CNRS through the MITI interdisciplinary programs and by IRSN through its exploratory research program. The project was also made possible thanks to the credits of the Hubert Curien Maimonides program made available by the French Ministry of Europe and Foreign Affairs and the French Ministry of Higher Education and Research. We acknowledge the financial support of the IdEx University of Bordeaux/Grand Research Program “GPR LIGHT”. This work was also supported by the OFFERR "NEXT" project under the European Union's Horizon 2020 research and innovation program and French Agence Nationale de la Recherche, under Project ANR LIOR (ANR-24-CE05-6070). We wish to acknowledge the support of the National Science Foundation (NSF Grant No. PHY-2206777) and the Czech Science Foundation (GA ČR) for funding on Project No. 22-42890L in the frame of the National Science Foundation–Czech Science Foundation partnership.
I.P. acknowledge the support of the Pazy Foundation Grant No.435/2023, and the support of the Israeli Institute for Fusion Research. This research was funded by the Israeli Ministry of Innovation, Science \& Technology, and by the Israel Institute for the Study of Nuclear Fusion and Hot Plasma.
ELI-NP-affiliated authors acknowledge support by the Extreme Light Infrastructure-Nuclear Physics (ELI-NP) Phase II, a project co-financed by the Romanian Government and the European Union through the European Regional Development Fund, by the Romanian Ministry of Education and Research through the Projects Nucleu (Grant No. PN 23210105), ELI-RO/RDI/2024{\_}016 Delphi and ELI-RO/RDI/2024{\_}028 Flight. The Romanian Government supports ELI-NP user facility operation through the IOSIN program, funding ELI-NP as a Facility of National Interest.
\end{acknowledgments}

\section*{DATA AVAILABILITY}
The data that support the findings of this article are not
publicly available upon publication because it is not
technically feasible. The data are available from the
authors upon reasonable request.



\bibliography{main}

\end{document}